\documentclass{aa}
\usepackage{psfig}
\def\H0 {$H_{\rm o}$}

\def\numd {\hbox{$n({\rm H}_2$)}}                   

\def\kms {\hbox{${\rm km\,s}^{-1}$}}

\def\Kkms {\hbox{${\rm K\,km\,s}^{-1}$}}
\def\percc {$\hbox{{\rm cm}}^{-3}$}    %cm-3
\def\cmsq  {$\hbox{{\rm cm}}^{-2}$}    %cm-2
\def\ffs {\hbox{$.\,\!\!^{\rm s}$}}

\def\CH3C2H {\hbox{${\rm CH}_3{\rm C}_2{\rm H}$}} %CH3C2H
\def\greekg1 {(\zeta _{\rm i},\eta _{\rm j})}

\begin{document}

\thesaurus{ 03                         % Extragalactic astronomy
           (11.09.1 Maffei~2           % Galaxies: individual
            11.09.4                    % Galaxies: ISM
            13.19.1)}                  % Radio lines: galaxies  

\title{Dense gas in nearby galaxies
               }

\subtitle{XIV. Detection of hot ammonia in Maffei~2}

\author{Henkel, C.\inst{1}, Mauersberger, R.\inst{2}, Peck, A.B\inst{1}, 
Falcke, H.\inst{1}, Hagiwara, Y.\inst{1}
}

\offprints{C. Henkel}

\institute{
  Max-Planck-Institut f{\"u}r Radioastronomie,
  Auf dem H{\"u}gel 69, D-53121 Bonn, Germany
\and
  Instituto de Radioastronom\'{\i}a Milim{\'e}trica, Avda. Divina Pastora, 7NC,
  E-18012 Granada, Spain
}

\titlerunning{Extragalactic ammonia}

\authorrunning{C. Henkel, R. Mauersberger, A.B. Peck, H. Falcke, Y. Hagiwara
}

\date{Received date / Accepted date}

\maketitle

\markboth{C. Henkel, R. Mauersberger, A. Peck, H. Falcke, Y. Hagiwara: 
Extragalactic ammonia
}{}

\begin{abstract}

The ($J,K$) = (1,1), (2,2), (3,3), and (4,4) inversion lines of ammonia 
(NH$_3$) have been detected toward the nuclear 40$''$ sized bar of the 
nearby spiral galaxy Maffei 2. The relative intensities of the ammonia 
lines are characterized by a rotational temperature of 85\,K. This is higher 
than rotational temperatures measured toward IC\,342 and most Galactic Center 
clouds, implying kinetic temperatures $\ga$100\,K. Since the kinetic 
temperature of the gas is larger than that of the dust, NH$_{3}$ is tracing 
a particularly dense warm gas component that is heated by young massive stars, 
cloud-cloud collisions, or ion-slip heating in the nuclear starburst. The gas 
north of the nucleus ($V_{\rm LSR}$ = --80\,\kms) is more highly excited than 
the gas further south (+6\,\kms). This asymmetry might be related to pronounced
morphological distortions that are observed in the north-eastern part of the 
galaxy.

\ \ \ \ \\


\end{abstract}

\section{Introduction} 

For the past decade, it has been possible to study the excitation of the 
molecular gas in external galaxies (e.g. Mauersberger \& Henkel 1989; 
Mauersberger et al. 1990; Jackson et al. 1995; H{\"u}ttemeister et al. 
1997; Paglione et al. 1997; Heikkil{\"a} et al. 1999; Mao et al. 2000). 
While the interpretation of CO emission often requires the application 
of complex models that incorporate temperature gradients at the surfaces 
of clouds, species like CS, HCN, or HC$_3$N trace higher density gas 
closer to the cloud cores and may thus be properly modeled with the 
standard `Large Velocity Gradient' approach. Observed line intensity 
ratios of such high density tracers depend mostly on the number density 
\numd, but also on the kinetic temperature $T_{\rm kin}$ of the gas. It 
is virtually impossible to disentangle these parameters, making observations
of a molecule tracing exclusively temperature highly desirable.

Ammonia (NH$_3$) is such a molecule. With a single telescope-receiver 
configuration at 18--26\,GHz, a large number of NH$_3$ inversion transitions 
can be measured, covering an enormous range of molecular excitation levels. 
These include the ($J,K$) = (1,1) to (4,4) inversion lines at 23, 65, 123, 
and 199\,K above the ground state. Extragalactic ammonia was first detected 
in the ($J,K$) = (1,1) line toward the nearby face-on spiral IC\,342 (Martin 
\& Ho 1979). This was followed by detections of the (2,2), (3,3), and
(4,4) lines (Ho et al. 1982, 1990; Martin \& Ho 1986) that permitted a first 
estimate of the kinetic temperature of the dense nuclear gas in an external 
galaxy. 

In view of the four ammonia lines observed toward IC\,342 (the (6,6) line
was also tentatively detected) it is surprising how little is known about 
ammonia in other external galaxies. Motivated by our unpublished detections 
of ($J,K$) = (1,1) and (2,2) emission in 1994 and by recent improvements in 
receiver sensitivity, baseline stability, and accessible bandwidth, we present 
a multilevel ammonia study of Maffei 2, an optically obscured barred spiral 
galaxy that is located behind the plane of the Milky Way at a distance of 
$\sim$2.5\,Mpc (e.g. Karachentsev et al. 1997). High lower limits to the 
kinetic temperature of its dense nuclear gas are derived.

\section{Observations}

All data presented here were taken in March 2000, using the Effelsberg 100-m 
telescope of the MPIfR equipped with a dual channel K-band HEMT receiver. The 
system temperature was $\sim$220\,K on a main beam brightness temperature 
scale; the beam size was $\sim$40$''$. The data were recorded using an 
autocorrelator with 8$\times$256 channels and bandwidths of 80\,MHz. The eight 
backends were configured in two groups of four, sampling data from both linear 
polarizations. Frequency shifts between the four backends representing a given 
linear polarization were adjusted in such a way that the ($J,K$) = (1,1) to 
(4,4) lines of ammonia could be measured simultaneously.
 
The measurements were carried out in a dual beam switching mode (switching
frequency 1\,Hz) with a beam throw of 121$''$ in azimuth. Calibration was 
obtained by measurements of the radio continuum of W3(OH) and 3C\,286 (for 
fluxes, see Baars et al. 1977; Mauersberger et al. 1988; Ott et al. 1994) 
and NH$_3$ spectra of Orion-KL and IC\,342 (Martin \& Ho 1986; Hermsen et 
al. 1988). We estimate an absolute calibration accuracy of $\pm$10\%. Relative 
calibration should be even more accurate, at the order of $\pm$5\%. Pointing 
errors as well as weather or elevation dependent gain variations affect all 
lines in a similar way; the calibration signal injected by a noise diode shows 
little frequency dependence and remains stable (within at least 5\%) over 
the observed frequency interval. Spectral intensities were converted to a 
main beam brightness temperature ($T_{\rm mb}$) scale. Pointing was checked 
every hour on nearby continuum sources and was found to be stable to within 
5--10$''$.

\begin{figure}
\hspace{1.3cm}
\psfig{figure=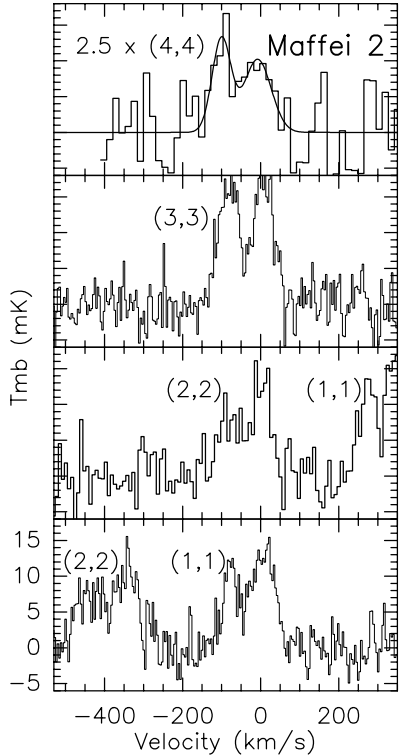,height=12.0cm,angle=-90}
\vspace{-1.3cm}
\caption[]{NH$_3$ spectra towards Maffei 2 ($\alpha_{2000}$ = 
2$^{\rm h}$ 41$^{\rm m}$ 55\ffs1, $\delta_{2000}$ = +59$^{\circ}$ 
36$'$ 12$''$) after subtraction of linear baselines. Channel spacings 
are 3.9, 7.8, 3.9, and 15.7\,\kms\ for the $(J,K)$ = (1,1) to (4,4) 
lines at 23694.496, 23722.631, 23870.130, and 24139.417\,MHz, respectively. 
The NH$_3$ (4,4) spectrum, with the temperature scale multiplied by a factor 
of 2.5, is shown with a tentative gaussian decomposition. }
\label{fig:spectra}
\end{figure}

\section{Results}

Fig.\,\ref{fig:spectra} shows the measured spectra toward the central 0.5\,kpc,
displaying a velocity range between --530 and +350\,\kms. Line parameters are 
given in Table \ref{tab:intensities}. The (1,1) to (3,3) lines are clearly 
detected; the (4,4) line is observed at a 4.5$\sigma$ level. The (1,1) to 
(3,3) profiles show two distinct velocity components, separated by 
$\sim$85\,\kms. In the (1,1) and (2,2) lines, the higher velocity component 
is stronger; in the (3,3) line, however, both components show similar line 
strength. In the (4,4) transition it is difficult to resolve the two features 
but the line center is closer to the velocity of the --80\,\kms\ component 
(Table 1). This is also suggested by a tentative two component fit to the 
(4,4) line shown in Fig.\,\ref{fig:spectra}, where 60--65\% of the emission 
seems to arise from the {\hbox{--80\,\kms}} component. We conclude that there 
is a trend with the --80\,\kms\ component rising in relative strength with 
excitation above the ground state. 

\begin{table}
\caption[]{\label{tab:intensities} Integrated line intensities 
($\int{T_{\rm mb} {\rm d}v}$), Local Standard of Rest velocities 
($V_{\rm LSR}$), full width to half power linewidths ($\Delta V_{1/2}$),
and upper state column densities (2$\cdot$$N_{\rm u}(J,K)$ $\sim$ $N$($J,K)$)
for the (1,1) to (4,4) lines of ammonia towards Maffei 2. Line parameters 
were obtained from gaussian fits to the data. Errors in Cols.\,3 and 4 
are standard deviations from the fit; Cols.\,2 and 5 also include a 10\% 
uncertainty in absolute calibration (see Sect.\,2).}
\begin{flushleft}
\begin{tabular}{ccccc}
\hline
Transition &$\int{T_{\rm mb}\,{\rm d}v}$& $V_{\rm LSR}$ & $\Delta V_{1/2}$ &
            $N_{\rm u}(J,K)$\\
\multicolumn{1}{c}{($J,K$)} & \multicolumn{1}{c}{\Kkms} & 
                              \multicolumn{2}{c}{\kms}  &
                              \multicolumn{1}{c}{10$^{12}$\,\cmsq} \\
\hline
     &               &           &                        \\
(1,1)&0.527$\pm$0.080&--78$\pm$2 & 43$\pm$6 &3.44$\pm$0.52\\
(1,1)&1.024$\pm$0.121&  +6$\pm$2 & 70$\pm$5 &6.69$\pm$0.78\\
     &               &           &                        \\
(2,2)&0.694$\pm$0.122&--78$\pm$6 & 83$\pm$12&3.40$\pm$0.60\\
(2,2)&0.782$\pm$0.131&  +5$\pm$4 & 66$\pm$10&3.83$\pm$0.63\\
     &               &           &                        \\
(3,3)&1.127$\pm$0.131&--82$\pm$2 & 62$\pm$4 &4.88$\pm$0.57\\
(3,3)&1.110$\pm$0.128&  +8$\pm$2 & 56$\pm$4 &4.80$\pm$0.55\\
     &               &           &                        \\
(4,4)&0.882$\pm$0.215&--57$\pm$20&174$\pm$40&3.55$\pm$0.87\\
\hline
\end{tabular}
\end{flushleft}
\end{table}
 
\section{Determination of rotational temperatures}

With h$\nu$/k $\sim$ 1.14\,K and expected excitation temperatures 
$\ga$10\,K across an inversion doublet we obtain, in the optically thin 
case, beam averaged column densities of 
$$
N(J,K) = \frac{7.77 \cdot 10^{13}}{\nu}\,\,\frac{J(J+1)}{K^2}\,\,
         \int{T_{\rm mb}\,{\rm d}}v
$$
for an individual inversion state. The column density $N$, the frequency 
$\nu$, and the integral are in units of cm$^{-2}$, GHz, and \Kkms, 
respectively. Calculated column densities are displayed in the last 
column of Table 1. For our observed metastable ($J$ = $K$) inversion 
transitions, the rotational temperature $T_{\rm rot}$ between the states 
($J,J$) and ($J',J'$) can then be obtained from 
$$
\frac{N(J,J)}{N(J',J')} = \frac{g_{\rm op}(J')}{g_{\rm op}(J)}\,\,
          \frac{2J' + 1}{2J + 1}\,\,\,
          {\rm exp}\left(\frac{-\Delta E}{{\rm k}\,T_{\rm rot}}\right),
$$
where $g_{\rm op}$ = 1 for para-ammonia, i.e. for ($J,K$) = (1,1), (2,2),
and (4,4), and $g_{\rm op}$ = 2 for ortho-ammonia, i.e. for the (3,3) line.
$\Delta E$ is the energy difference between the states involved. 

\begin{figure}
\hspace{0.5cm}
\psfig{figure=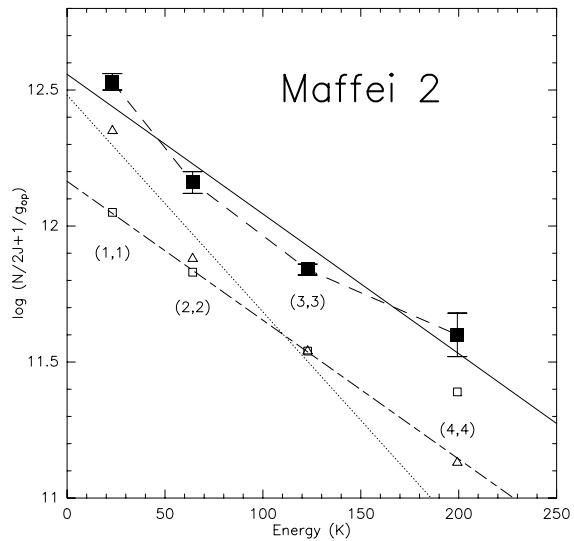,height=7.6cm,angle=-90}
\caption[]{A rotation diagram of metastable ammonia transitions towards 
Maffei 2, assuming optically thin emission and excitation temperatures 
$\gg$2.7\,K for each inversion doublet. The column densities $N$ from
Table 1 refer to the upper state of the respective inversion doublet. 
Filled squares: Total emission (dashed line connects the points); empty 
squares: --80\,\kms\ component; empty triangles: +6\,\kms\ component. 
Straight lines provide best fits to the total emission in the ($J,K$) = 
(1,1) -- (4,4) lines (solid) and to the (1,1) -- (3,3) emission lines for 
the --80 and +6\,\kms\ components (dashed and dotted, respectively). The 
decomposition of the (4,4) line into two velocity components, based on 
the fit shown in Fig.\,\ref{fig:spectra}, is at best only marginally 
significant.  Errors for the individual (1,1) to (3,3) components 
are $\sim$40\% larger than those for the total emission.}
\label{fig:trot}
\end{figure}

Rotation diagrams of the ammonia emission in the (1,1) to (4,4) lines 
are shown in Fig.\,\ref{fig:trot}. The (3,3) line, although belonging 
to a different ammonia species, fits well in the general trend of 
decreasing normalized column density with energy above the ground 
state. The slope of the lines fitting the normalized column densities 
is {\hbox{--log\,e/$T_{\rm rot}$}}. A weighted fit to the total ammonia 
emission (solid line; relative weights reflect signal-to-noise ratios) yields 
a rotational temperature of 85$\pm$15\,K (1$\sigma$ error). Excluding the 
(4,4) line from the fit gives 65$\pm$10\,K. For the two velocity components, 
only including the (1,1) to (3,3) lines, $T_{\rm rot}$ = 85$\pm$5 
($V_{\rm LSR}$ = --80\,\kms) and 55$\pm$10\,K ($V_{\rm LSR}$ = +6\,\kms). 
Inclusion of the highly tentative decomposition of the ($J,K$) = (4,4) line 
(Fig.\,\ref{fig:spectra}) yields $T_{\rm rot}$ $\sim$ 115 and 65\,K. 
As sometimes observed in Galactic sources, the inclusion of higher excited 
inversion levels leads to higher rotational temperature estimates. This is 
indicated by the decreasing (negative) slope of the line connecting 
the measured points. 

\section{Discussion}

\subsection{Interpretation of rotational temperatures}

The $J = K$ ammonia inversion doublets are called metastable because
they decay radiatively via slow $\Delta K$ = $\pm$3 transitions. 
As a consequence, the relative population of these states is strongly 
affected by collisions and thus reflects (e.g. Walmsley \& Ungerechts 
1983) the kinetic temperature of the gas ($T_{\rm rot}$ $\leq$ 
$T_{\rm kin}$).

One way to explain rising rotational temperatures with increasing energy 
above the ground state is the presence of kinetic temperature gradients. 
Alternatively, optically thick lines or subthermal excitation can produce 
the same effect (Goldsmith \& Langer 1999). Accounting for these latter 
two effects, radiative transfer calculations (e.g. Walmsley \& Ungerechts 
1983; Schilke 1989) predict that rotational temperatures approach the 
kinetic temperature with rising energy of the analysed metastable inversion 
doublets. This yields rising $T_{\rm rot}$ with rising $E$/k. It is therefore
likely, based on the corrections of the models, that the true kinetic
temperature is $>$100\,K. $T_{\rm kin}$ should be larger for the --80 than 
for the +6\,\kms\ velocity component. 

\subsection{A comparison with IC\,342 and Galactic Center clouds}

NH$_3$ integrated line temperatures are {\it larger} towards Maffei 2 than 
towards IC\,342, also located at $D$$\sim$2.5\,Mpc (Karachentsev et al. 1997).
Emission from the inner 500\,pc of the Milky Way would show similar
intensities if observed from this distance (e.g. H{\"u}ttemeister et al. 
1993a,b). 

From the ($J,K$) = (1,1) to (4,4) lines, $T_{\rm rot}$$\sim$85\,K in Maffei 
2, but only 50$\pm$10\,K in IC\,342 (Martin \& Ho 1986). Toward peak positions 
of clouds near the center of the Galaxy, (1,1) lines tend to be optically 
thick, while more highly excited inversion lines show no significant optical 
depth effects (H{\"u}ttemeister et al. 1993b). There are two main temperature 
components, one at $T_{\rm kin}$ $\sim$ 25 and one at $\sim$200\,K. These, if 
present in Maffei 2 as well, could be responsible for the increasing rotational 
temperatures with $E$/k (Fig.\,\ref{fig:trot}). Such gradients are also seen 
in IC\,342, if the (6,6) line is as strong as suggested by the spectrum
presented by Martin \& Ho (1986). 

\subsection{The nuclear region of Maffei 2}

What kind of molecular gas is seen in ammonia? Critical densities, where
collisional de-excitation matches spontaneous emission rates, are \numd\ $\sim$
3$\cdot$10$^{3}$\percc\ (e.g. Green 1980; Ho \& Townes 1983). Our beam size 
(40$''$) matches the central bar with its size of 43$''$$\times$9$''$ in 
$^{12}$CO {\hbox{$J$ = 1--0}} (Ishiguro et al. 1989) and 40$''$$\times$12$''$ 
in $^{13}$CO $J$ = 1--0 (Hurt \& Turner 1991). The presence of bars 
provides a viable mechanism to transfer interstellar gas into the nuclear 
region of a galaxy, leading to the formation of dense molecular clouds 
and triggering bursts of star formation. As long as the gas is not 
yet fully ionized or swept away by stellar winds, it will coexist with the 
newly formed stars and will be heated by their radiation, by an enhanced 
flux of cosmic rays, by cloud-cloud collisions, and by the dissipation of 
turbulent energy. Both Maffei 2 and IC\,342 show prominent nuclear bars and 
strong CO, infrared, and radio continuum emission (e.g. Lo et al. 1984; 
Ishiguro et al. 1989; Ishizuki et al. 1990; Hurt \& Turner 1991; Turner 
\& Ho 1994; Meier et al. 2000). 

Assuming a grain emissivity law of $\lambda^{-1}$, dust temperatures 
deduced from the Kuiper Airborne Observatory and, for a larger region,
from IRAS data (Rickard \& Harvey 1983; IRAS 1989) are $T_{\rm d}$ =
33--47\,K. Ignoring non-metastable inversion doublets, Table 1 implies
that $N$(NH$_3$) is on the order of 10$^{14}$\,\cmsq. For H$_2$, 
with $D$=2.5\,Mpc, $N$(H$_2$)$\sim$10$^{22}$\,\cmsq\ (Hurt \& Turner 1991). 
This yields a relative abundance of $X$(NH$_3$)$\sim$10$^{-8}$. With
$T_{\rm rot}$(NH$_3$) $\gg$ $T_{\rm d}$ and an unusually small NH$_3$
abundance (e.g. Benson \& Myers 1983; Brown et al. 1988), ammonia should 
trace a gas component that is different from that seen in CO (for an 
analysis of CO emission in a nuclear starburst environment, see Mao et 
al. 2000). Cloud-cloud collisions, young massive stars, or ion-slip
heating (H{\"u}ttemeister et al. 1993a,b) may cause the high rotational
temperatures. SiO, an exclusive tracer of high temperature gas, was
detected in Maffei 2 (Sage \& Ziurys 1995) and is likely arising from the
same gas component. Detailed studies of NH$_3$ and SiO in the Galactic 
Center region (e.g. H{\"u}ttemeister et al. 1993a,b, 1998) indicate that 
the warm gas not directly associated with massive stars is only moderately 
dense (\numd $\sim$5$\cdot$10$^{3}$\percc).

From a comparison with the interferometric maps of Ishiguro et al. (1989) 
and Hurt \& Turner (1991), the ammonia component at +6\,kms\ must arise SW, 
the {\hbox{--80\,\kms}} component NE of the center of the galaxy. Why is the 
rotation diagram (Fig.\,\ref{fig:trot}) indicating higher excitation NE of 
the nucleus? The bar in Maffei 2 may be caused by a merger with a satellite 
galaxy in the NE that severely disrupts this side of the galaxy (Hurt et al. 
1996). This might have led to differences in inflow rates, star forming rates, 
and gas heating at the north-eastern and south-western sides of the nucleus. 
While our observed small scale ($\sim$500\,pc) nuclear asymmetry in rotational 
temperature may be consistent with the large scale asymmetry observed in H{\sc 
i}, the proposed connection between these phenomena remains speculative. More 
data on the putative interacting dwarf galaxy, detailed simulations of the 
interaction, and models of the evolution of the nuclear bar are needed to 
verify or to reject the proposed scenario and to explain the recent past of 
Maffei 2.

\begin{acknowledgement} 

We wish to thank J.L. Turner for critically reading the manuscript.

\end{acknowledgement}

\end{document}